# Below band gap formation of solvated electrons in neutral water clusters?

*Loren Ban, Christopher W. West, Egor Chasovskikh, Thomas E. Gartmann, Bruce L. Yoder and Ruth Signorell\**

ETH Zurich, Department of Chemistry and Applied Biosciences, Vladimir-Prelog-Weg 2, CH-8093 Zurich, Switzerland

**ABSTRACT**: Below band gap formation of solvated electrons in neutral water clusters using pump-probe photoelectron imaging is compared with recent data for liquid water and with above band gap excitation studies in the liquid and clusters. Similar relaxation times in the order of 200 fs and 1-2 ps are retrieved for below and above band gap excitation, in both clusters and liquid. The relaxation times independence from the generation process indicates that these times are dominated by the solvent response, which is significantly slower than the different solvated electron formation processes. The analysis of the temporal evolution of the vertical electron binding energy and the electron binding energy at half maximum suggests a dependence of the solvation time on the binding energy.

**TOC GRAPHICS**

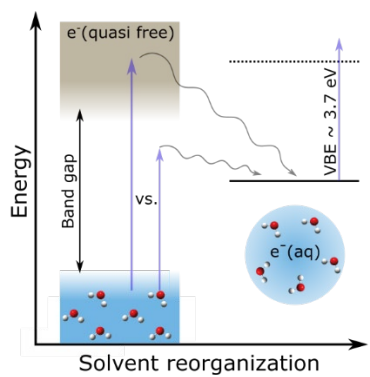

**KEYWORDS**: photoelectron spectroscopy, hydrated electron, electron scattering, electron binding energy, ultrafast imaging.

**Introduction**

Solvated electrons in molecular liquids, especially water, have sparked broad interest because of their widespread occurrence and their intriguing fundamental properties. Numerous experimental and theoretical studies have provided insights into the electronic properties, the generation mechanism and the relaxation dynamics of solvated electrons in liquid water, amorphous ice, anionic water clusters, neutral water clusters and sodium-doped water clusters[1–48] (and references therein).

A series of comparable, femtosecond time-resolved photoelectron studies on liquid water and neutral water clusters[49–51] has now opened up the possibility to address two aspects in more detail: (i) Similarities and differences of the relaxation dynamics after below versus above band gap generation of the hydrated electron, and (ii) Similarities and differences of the relaxation dynamics in two different aqueous environments, i.e. liquid water versus neutral water clusters. The formation of hydrated electrons by irradiation with photons has been observed down to the absorption edge of water of ~6 eV[37–39,52] (and references therein), far below the bottom of the conduction band of water[53–56]. Different mechanisms have been suggested for the below band gap formation of solvated electrons, involving very fast (<10 fs) water dissociation and proton and electron transfer processes ("hot H atom mechanism", "proton-coupled electron transfer", "consecutive proton transfer, electron transfer process", "electron transfer to preexisting sites"; see refs.[37–39,52] and references therein). These ultrafast processes are followed by slower solvent rearrangement (fs to ps timescales) and slow geminate recombination[37,40,47,57–59] (and references therein). Irradiation with photon energies above the band gap, by contrast, produces delocalized, quasi-free conduction band electrons by ionization of water. These conduction band electrons relax rapidly towards the bottom of the conduction band by fast energy dissipation through electron scattering on timescales of several 10 fs[15,31,60–63]. Localization from the bottom of the conduction band to inter-band trapped states likely takes place on timescales faster than ~100 fs[15,25,30,34], followed by slower solvent rearrangement and geminate recombination.

The present time-resolved photoelectron imaging study addresses the question of whether the formation of solvated electrons by below band gap excitation is also feasible in neutral water clusters, as previously observed in liquid water[38,50] (and references therein). The relaxation dynamics after one-photon excitation at 7.8 eV photon energy is probed and compared with a recent pump-probe study for liquid water using an excitation energy of 7.7 eV[50]. At these excitation energies, it is assumed that solvated electron formation follows absorption into a localized $1^1B_1$ excited state of water. The results for below band gap excitation are compared with previous above band gap excitation photoelectron studies for the liquid and clusters[49–51].

**Experimental Details**

The velocity map imaging (VMI) photoelectron spectrometer used in this work is described in refs.[36,49,64,65] and the supporting information (SI). Neutral water clusters consisting of ~500 water molecules ($r$~1.5 nm, see refs.[64,66] for cluster size determination) are generated in supersonic expansions. Femtosecond (fs) laser pulses of 7.8 eV photon energy (pump) from high harmonic generation (HHG) are used to generate solvated electrons by below band gap excitation in water clusters. The relaxation dynamics are probed with a femtosecond probe pulse of a 4.7 eV photon energy by varying the pump-probe time delay $t$. Time-resolved electron binding energy spectra (eBE) and photoelectron angular distributions (PADs, $\beta$-parameters Eq. S1, SI) are retrieved from the recorded time-dependent photoelectron images (SI). The instrument response function (IRF) determined from (1+1') non-resonant ionization of Xe is $150 \pm 35$ fs.

**Results**

Fig. 1a shows time-resolved photoelectron spectra (TRPES) recorded with a 4.7 eV probe pulse following below band gap excitation at 7.8 eV pump energy. Within the first picosecond, the eBE spectrum changes rapidly and approaches an asymptote at longer times, accompanied by a decrease in the overall signal intensity. The observed spectral signature is characteristic for hydrated electron relaxation dynamics, where the fully relaxed hydrated electron in its ground state is formed after ~2 ps[49–51], followed by depopulation due to geminate recombination on a much longer timescale[50]. This confirms hydrated electron formation in neutral clusters following below band gap excitation.

We expect the short-lived signal observed around zero time delay extending towards maximum eBEs of ~4.7 eV (towards zero photoelectron kinetic energy, eKE) to originate either from the fast dynamics (much faster than our IRF) of the background water vapour[67] or the impulsive (1+1') ionization of water clusters. To subtract these spectral components, we performed a Global Lifetime Analysis (GLA) of our data[18,68,69]. GLA assumes a number of time-independent spectra and assigns them a corresponding exponentially decaying population. The spectral profiles and the decay dynamics are then fit to the experimental data. This technique has recently been applied by Hara *et al.*[69] to model detailed kinetics following conduction-band excitation of methanol. Since the relaxation dynamics are faster in water than in methanol[50,69,70], much better time resolution than currently available and excellent signal-to-noise levels would be required to extract similarly detailed kinetic information for aqueous systems. We thus use here a simple sequential kinetics model (Eq. S2, SI) with three exponentially decaying features and a Gaussian distribution for the IRF to fit the TRPES in Fig. 1a (SI, section S2), similar to the approaches in refs.[50,69]. The best TRPES fit (Fig. S1, middle panel) corresponds to a Gaussian with a full-width at half-maximum FWHM$_{\text{IRF}}$ = 166±5 fs, consistent with the experimentally determined IRF of 150 ± 35 fs, and three exponential decays with lifetimes of $\tau_1^{\text{GLA}}$ = 221±56 fs, $\tau_2^{\text{GLA}}$ = 447±100 fs and $\tau_3^{\text{GLA}}$ = 17.74±5.55 ps (SI and Fig. 2) with the corresponding Decay Associated Spectra (DAS; Fig. 2b). This analysis allows us to subtract the impulsive component from the TRPES (Fig. 1b, see also eBE spectra in Figs. S3 and S4).

Fig. 2a shows the time-dependent, relative photoelectron yield of the hydrated electron (experimental data: circles, GLA fit: full black line) with the contributions from the corresponding decay components (dashed colored lines). The hydrated electron yield decays to ~45% after 2 ps and to ~ 33% after 5 ps of the yield measured at 200 fs. Within our uncertainties (±15%), these values are comparable to the measurements in liquid water of ~31% and ~22%, respectively[50], hinting at similar loss mechanisms in clusters and liquid, at least for these relatively short pump-probe delay range. Electron yields recorded at long pump-probe time delays have previously been used to extract geminate recombination rates, survival probabilities and retrieved electron ejection lengths in the liquid[50] (and references therein). In clusters, it is difficult to obtain reliable experimental data for long pump-probe delays because of the generally lower signal to noise level.

Experimental time-dependent eBE spectra are shown before subtraction of the impulsive component in Fig. S2 and after subtraction of the impulsive component in Fig. S3. Fitting these spectra with an exponentially modified Gaussian function (Eq. S6 and Figs. S2 and S3) allows us to retrieve the time evolution of the vertical electron binding energy (VBE; most probable eBE) and the electron binding energy at half maximum (HBE, see below). At short time delays <60 fs, subtraction of the impulsive component is required to obtain reliable VBEs, while the VBEs at longer time delays are indistinguishable within our experimental uncertainty regardless of whether the impulsive component is subtracted or not (Fig. S2 and S3). The experimental VBEs at time delays >60 fs are shown in Fig. 3a (circles). The VBE shifts from an initial value of 3.0 eV to 3.75 eV, which is reached at ~2 ps and remains constant afterwards. The eBE spectra with a VBE of 3.75 eV recorded after ~2 ps (Figs. S2 and S3) coincide with those of the relaxed ground state hydrated electron recorded in clusters and the liquid at a probe energy of 4.7 eV[2,13,31,46,49,71]. We have also performed cluster size-dependent studies for average cluster sizes between <n> ~250 and 500 molecules. No cluster size-dependence of the VBE (at ~2 ps) is observed within the experimental uncertainty (see ref.[49] for comparison with water anion and Na-doped water clusters). In agreement with our previous study using 10.9 eV pump photons[49], we retrieve a $\beta$-parameter of ~0.2 for the relaxed ground state hydrated electron (after ~2 ps). After scattering corrections[31,49,61,64], this corresponds to a genuine $\beta$-parameter in the range of 0.51-0.66. For the present pump photons of 7.8 eV, no significant time-dependence is observed for the $\beta$-parameter, i.e. the value of $\beta$ lies around ~0.2 for all pump-probe delays. This contrasts with our previous results using 10.9 eV pump photons, where the initial ($t$ = 0 fs) $\beta$-parameter of ~0.4 exceeded the ground state value of ~0.2[49]. However, such differences must be viewed with caution given our relatively high experimental uncertainty of ~±0.15.

**Discussion**

The temporal evolution of the VBE provides information on the relaxation dynamics following below band gap excitation and allows us to compare them with data from previous studies. We find that the time-dependence of the VBEs in Fig. 3a (black circles) is well represented by a biexponential function (black full line): $VBE(t) = a_1 \exp(-t/\tau_1^{VBE}) + a_2 \exp(-t/\tau_2^{VBE}) + VBE(t=\infty)$, with VBE(t=∞)=3.8 eV and time constants $\tau_1^{VBE}$ = 220 fs and $\tau_2^{VBE}$ = 1.6 ps. These time constants agree well (see Table 1) with those reported in a recent liquid microjet study of water by Suzuki and coworkers[50], following below band gap excitation using a very similar pump energy of 7.7 eV (blue diamonds in Fig. 3a). A similarly good agreement between the relaxation times of clusters and liquid was also observed at a pump energy of 9.3 eV (Table 1). Considering that $\tau_1^{VBE}$ and $\tau_2^{VBE}$ are likely dominated by the response of the solvent (see introduction and ref.[15]), the similarity of clusters and liquid does not appear so obvious. Clusters have more surface and are thus structurally less restricted than the liquid. One would thus expect solvent rearrangement to be faster in the cluster than the liquid. On the other hand, the cluster temperatures are likely lower than that of the liquid, which would result in the opposite effect on the relaxation times, i.e. one would expect an increase of the solvent rearrangement times in the cluster compared with the liquid. Furthermore, one would expect both effects to be more pronounced for the slower solvent motions (larger amplitude reorientation), i.e. more pronounced for $\tau_2^{VBE}$ than for $\tau_1^{VBE}$. This appears to be the case for the data at 7.8 and 9.3 eV (see Table 1). The slightly higher values of $\tau_2^{VBE}$ in the clusters might imply that the cluster temperature is the dominant effect. However, such a conclusion remains speculative with the caveat of the limited comparability of cluster data with the liquid jet results. Apart from the somewhat different analysis methods used, $\tau_2^{VBE}$ is generally less well constrained by the cluster data with their lower signal-to-noise and fewer data points at longer time delays.

The observed independence of $\tau_1^{VBE}$ and $\tau_2^{VBE}$ on the pump energy used (see data for the different $h\nu$ in Table 1) supports the conclusions that these timescales are dominated by solvent response. The generation process will mainly influence the very fast processes (e.g. electron scattering, localization from the conduction band into the band gap, the different below band gap formation mechanisms) taking place on timescales of a few tens of fs, while the generally slower solvent responses will at most weakly depend on the generation process. The fast component of the solvent relaxation would still be expected to correlate to some extent with electronic deactivation and localization, so that it might come as a surprise to find almost identical values for $\tau_1^{VBE}$ at all different pump energies (Table 1). However, one needs to keep in mind that these values lie close to the time resolution (on the order of ~100 fs) of the studies. To properly access these fast processes, one would need a significantly better time resolution by about a factor of ~10. Observable effects of using different excitation energies are mainly limited to the PADs and VBEs at zero pump-probe delay ($t$~0 fs). As mentioned above, the initial $\beta$-parameter recorded at 10.9 eV appears to be higher (~0.4) than the one measured at 7.8 eV (~0.2). Similarly, the initial VBE (VBE($t$~0 fs)) seems to decrease systematically with increasing pump energy by about 0.7 eV (Fig. 3b). At 7.8 eV, VBE($t$~0 fs) lies ~0.7 eV below the values for the relaxed ground state of the hydrated electron (horizontal dashed line at ~3.75 eV), while this difference increases to ~1.4 eV at 15.5 eV pump energy. Together with the above, this suggests that fast electronic relaxation processes remain largely hidden within the time resolution. As a result, different pump-probe studies in Table 1 probe very similar solvation dynamics probably starting from similar localized initial states in the band gap, independently of the pump energy.

The temporal evolution of the VBE is a simple representation of the complex underlying dynamics because it only probes the dynamics of states that correspond to the maximum of the eBE spectrum (Fig. S2). In Fig. 3a, we also show the time-dependence of the binding energy at half maximum (HBE, red triangles) on the rising edge of the eBE spectrum together with a biexponential fit (red dashed line), resulting in time constants of $\tau_1^{HBE}$ = 220 fs and $\tau_2^{HBE}$ = 3.3 ps. $\tau_2^{HBE}$ exceeds $\tau_2^{VBE}$ by a factor of two. The HBE data represents the temporal evolution of a population with a lower binding energy compared with the population represented by the VBE data. This implies a correlation between the binding energy and the relaxation time. A potential explanation for this correlation could be that states with lower binding energy require more time to relax because more extensive solvent rearrangement is needed to reach the relaxed ground state. To represent this behavior, we therefore suggest using the HBE in addition to the VBE. A comparison with previous studies is currently not possible because the temporal evolution of the HBE is not available from those studies. Finally, we would like to highlight another advantage of the HBE when comparing different studies. Many of the solvated electron photoelectron studies use ultraviolet probe photons in the range below 5.8 eV; i.e. in a range where the energy dependence of electron scattering has a pronounced influence on the shape of the measured binding energy spectra[2,31]. Fig. 3 in ref.[31] shows that the position of the HBE only shifts by 50 meV when changing the probe photons from 4.4 to 5.8 eV, while the position of the VBE shifts by more than 200 meV over the same probe energy range. The HBE is less affected by the influence of scattering than the VBE, and thus makes comparisons between studies using different probe wavelengths more reliable. We would like to emphasize that for solvated electrons in water the preferred method to correct for such biases arising from electron scattering are detailed scattering simulations to extract genuine binding energy spectra [31,61–64,72]. However, in most studies this correction has not yet been implemented.

**Conclusion**

We have investigated the relaxation dynamics of the solvated electron in neutral water clusters (~500 molecules) after below band gap excitation with photons of 7.8 eV using ultrafast photoelectron imaging. The comparison with a recent liquid water study at an excitation energy of 7.7 eV[50] reveals similar solvation times on the order of ~200 fs and ~1-2 ps in the clusters and the liquid. It has been suggested that at these excitation energies, hydrated electrons are generated by water dissociation and proton and electron transfer processes[38] after one-photon absorption to the predominately localized $1^1B_1$ excited state in water. The comparison of the below band gap investigations with previous above band gap photoelectron studies[49–51] shows that the observable relaxation dynamics are essentially independent of the excitation energy. This is not surprising considering the timescales accessible in these experiments (~100 fs range) that mainly probe the solvent response, which is largely independent of the generation process of the solvated electron. A much higher time-resolution (by least better than a factor of 10) would be required to detect excitation energy dependent differences in the generation process of the solvated electron, arising from ultrafast processes, such as electron scattering, localization into the band gap, and different below band gap electron formation mechanisms. Finally, the different solvation dynamics observed for the binding energy at half maximum and the vertical electron binding energy suggest that states with a lower electron binding energy correlate with longer solvation times, potentially because more extensive solvent rearrangement is required to relax states with lower binding energy.

FIGURES

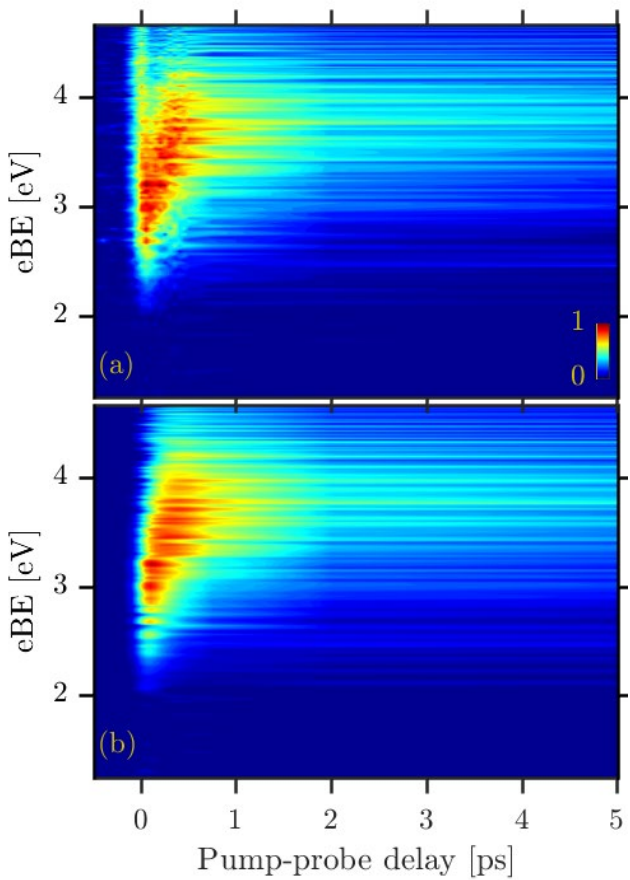

**Figure 1:** Time-resolved photoelectron spectra (TRPES) of the hydrated electron in neutral water clusters. **(a)** Experimental TRPES measured with a 7.8 eV pump pulse and a 4.7 eV probe pulse (without subtraction of the impulsive component). **(b)** Global Lifetime Analysis (GLA) fit to the TRPES in panel (a) after subtraction of the impulsive component at early pump-probe delays (see also Fig. S1).

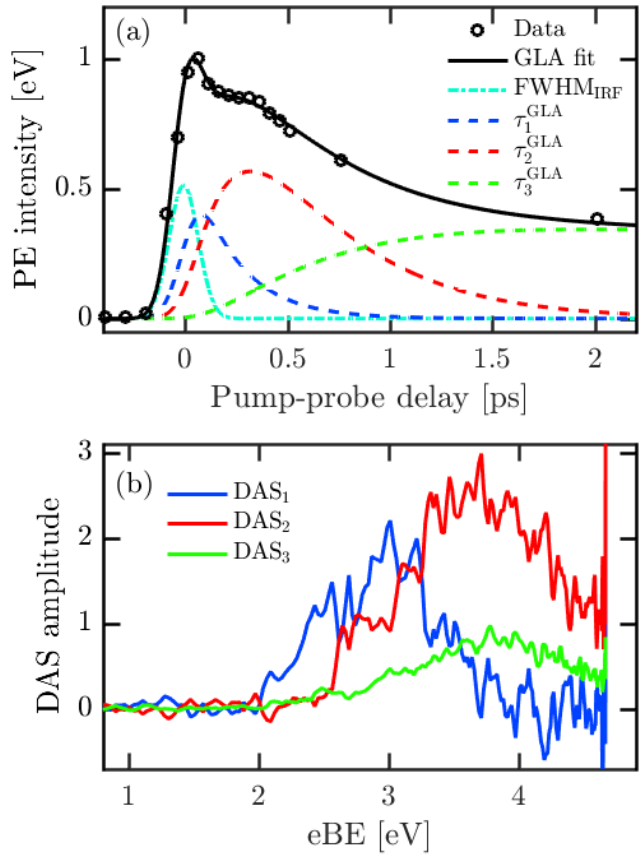

**Figure 2: (a)** Experimental (circles) and fitted (full black line) time-dependent relative photoelectron yield with the contributions from the corresponding decay associated spectra (DAS, dashed colored lines). The Gaussian component (cyan dashed-dotted line, $FWHM_{IRF}$ = 166±5 fs) matches well with the measured instrument response function. Contributions from $DAS_1$ (blue dashed line, $\tau_1^{GLA}$ = 221±56 fs), $DAS_2$ (red dashed, $\tau_2^{GLA}$ = 447±100 fs) line and $DAS_3$ (green dashed line, $\tau_3^{GLA}$ = 17.74±5.55 ps) follow a sequential kinetics mechanism. **(b)** The three DAS obtained from the fit (SI, section S2). $DAS_1$: blue line; $DAS_2$: red line and $DAS_3$: green line.

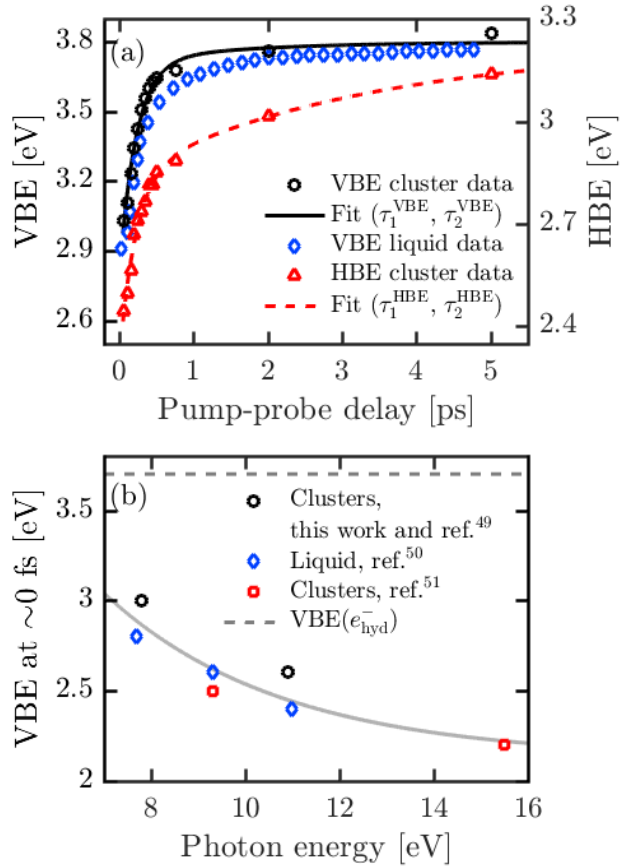

**Figure 3: (a)** Time-dependent vertical binding energy (VBEs, black circles; left abscissa) and binding energy at half maximum (HBEs, red triangles, right abscissa) for neutral water clusters recorded at a pump energy of 7.8 eV. The data are extracted from the TRPES in Fig. 1 (see text and SI). The biexpoential fit to the VBEs with time constants $\tau_1^{VBE}$ = 220 fs and $\tau_2^{VBE}$ = 1.6 ps is shown as the black full line. The biexpoential fit to the HBEs with time constants $\tau_1^{HBE}$ = 220 fs and $\tau_2^{HBE}$ = 3.3 ps is shown as the red dashed line. Blue diamonds: Time-dependent VBEs for liquid water extracted from Fig. 2d of ref.[50], recorded at a pump energy of 7.7eV. **(b)** VBE at a pump-probe delay of ~0fs (VBE(t~0fs), see Table 1) for the liquid from ref. [50] (blue triangles) and different cluster studies clusters (black circles: this work and ref.[49], red squares: ref[51]) as a function of the pump energy $h\nu$. The gray line is to guide the eye. The horizontal dashed line at ~3.75 eV indicates the VBE of the relaxed ground state hydrated electron ($e_{hyd}^-$) which is reached after ~2 ps. Note that the different studies use somewhat different analysis methods and different time resolutions.

**Table 1:** Comparison of the time constants $\tau_1^{VBE}$ and $\tau_2^{VBE}$ for the solvation of the hydrated electron in water clusters and liquid water retrieved from the time evolution of the VBEs. $h\nu$ is the pump photon energy. Except for the data at 11 eV all excitations are single-photon excitations. VBE (t~0fs) is the vertical binding energy at a pump-probe delay of ~0 fs. Note that the different studies use somewhat different analysis methods and different time resolutions.

| $h\nu$ [eV] | 7.8 and 7.7 | | 9.3 | | 10.9 and 11.0 | | 15.5 |
|---|---|---|---|---|---|---|---|
| | Cluster (this work) | Liquid[50] | Cluster[51] | Liquid[50] | Cluster[49] | Liquid (two-photon)[50] | Cluster[51] |
| $\tau_1^{VBE}$ [ps] | 0.2 | 0.2 | 0.21 | 0.3 | - | 0.2 | 0.18 |
| $\tau_2^{VBE}$ [ps] | 1.6 | 1.0 | 1.4 | 0.9 | - | 2.0 | 1.3 |
| VBE(t~0fs) [eV] | 3.0 | 2.8 | 2.5 | 2.6[a] | 2.6 | 2.4[a] | 2.2[b] |

[a]Values estimated from Figs. 2e and 2f in ref.[50].

[b]Value estimated from Fig. S6c in ref.[51].

## ASSOCIATED CONTENT

**Supporting Information**. The following files are available free of charge.

Additional details on the experiment and data analysis. (PDF)

## AUTHOR INFORMATION

**Corresponding Author**

*rsignorell@ethz.ch

**Notes**

The authors declare no competing financial interests.

## ACKNOWLEDGMENT

We thank David Stapfer and Markus Steger for technical support. This project has received funding from the European Union's Horizon 2020 research and innovation program from the European Research Council under the Grant Agreement No 786636, and the research was supported by the NCCR MUST, funded by the Swiss National Science Foundation (SNSF), through ETH-FAST, and through SNSF project no. 200020_172472. C.W.W. acknowledges funding from the European Union's Horizon 2020 research and innovation programme under the Marie Skłodowska-Curie grant agreement No 801459 - FP-RESOMUS - and the Swiss National Science Foundation through the NCCR MUST. R.S. is a grateful recipient of a Humboldt Research Prize form the Alexander von Humboldt Foundation and a Mildred Dresselhaus Guestprofessorship from the Centre for Ultrafast Imaging in Hamburg.

**Supplementary Material:**

# Below band gap formation of solvated electrons in neutral water clusters?

*Loren Ban, Christopher W. West, Egor Chasovskikh, Thomas E. Gartmann, Bruce L. Yoder and Ruth Signorell\**

ETH Zurich, Department of Chemistry and Applied Biosciences, Vladimir-Prelog-Weg 2, CH-8093 Zurich, Switzerland

AUTHOR INFORMATION

**Corresponding Author**

\*rsignorell@ethz.ch

# S1: Experimental

The experimental setup used in this work is described in refs.[1,2]. In short, the neutral water clusters are generated by supersonic expansion from a pulsed Even-Lavie valve[3] using Ne as a carrier gas. The sample tube with liquid water was heated to 384 K and the valve was kept at 388 K. Together with the Ne pressure of ~15 bar this yields a water:Ne mixture of 1:10.
Cluster size distributions are measured by the Na-doping method[4–9] using 4.7 eV light to ionize the doped clusters and ion extraction parallel to the cluster propagation direction for ion time-of-flight measurements[2]. The average cluster size used in this work is ~500 water molecules ($r \sim 1.5$ nm).

EUV light pulses used as a pump are generated by high harmonic generation (HHG)[10] in argon from a portion of the fundamental 1 kHz, 35 fs Ti:sapphire laser centered at 795 nm (Coherent Astrella). Light pulses of a single harmonic (5$^{th}$ harmonic, 7.8 eV) are selected in our time-preserving monochromator[11,12]. To probe the dynamics, we used 4.7 eV pulses generated by frequency up-conversion of the second fundamental beam portion in a pair of BBO crystals. The instrument response function (IRF) determined from (1+1') non-resonant ionization of Xe is 150 ± 35 fs. A constant background signal originating from Ne carrier gas is present in some of the measurements, originating from single-photon ionization (IP = 21.56 eV) by the 15$^{th}$ harmonic (23.4 eV) diffracted in the 3$^{rd}$ diffraction order at the grating position for 7.8 eV light. We expect no contribution from the 15$^{th}$ harmonic to the excitation process. The photon flux of the 3$^{rd}$ order diffracted 15$^{th}$ harmonic is 2-3 orders of magnitude lower than the 5$^{th}$ harmonic, but due to reasonably large cross sections[13–15] and the seeding ratio of 1:10 (1.5 bar H$_2$O, 15 bar Ne) the neon peak appears prominently in the spectrum. No co-clustering between neon and water was observed.

Photoelectron kinetic energies (eKEs) and angular distributions (PADs) are measured in the velocity map imaging (VMI) spectrometer[4,16]. Time-dependent electron binding energy (eBE) spectra are reconstructed from the velocity map images with MEVIR[17,18]. From the spectrometer resolution and day-to-day instabilities, we estimated an uncertainty in vertical binding energy determination of ~ 0.15 eV. The PADs are characterized by a single anisotropy parameter $\beta$[19,20] defined as

$$I(\theta) \propto 1 + \frac{\beta}{2}(3\cos^2\theta - 1), \quad (S1)$$

where $I(\theta)$ is photoelectron signal as a function of $\theta$, angle between the light polarization axis and ejection direction of the photoelectron.

## S2: Global Lifetime Analysis (GLA)

To fit the experimental TRPES (Fig. S1 top) we use the GLA approach with 3 exponentially decaying features and a single Gaussian feature representing the impulsive component. The exponentially decaying features follow a sequential kinetics model

$$A \xrightarrow{\frac{1}{\tau_1^{GLA}}} B \xrightarrow{\frac{1}{\tau_2^{GLA}}} C \xrightarrow{\frac{1}{\tau_3^{GLA}}}. \tag{S2}$$

The photoelectron binding energy (eBE) spectrum at time $t$ is given as

$$I(t) = S_{IRF}(t; eKE) + \sum_{i=1}^{3} S_i(t; eKE), \tag{S3}$$

where IRF is the instrument response function and

$$\begin{aligned}
S_{IRF}(t; eKE) &= DAS_{IRF} \exp\left[\frac{-t^2}{2\sigma_{IRF}^2}\right], \\
S_1(t; eKE) &= DAS_1 X_1 \exp(-k_1 t), \\
S_2(t; eKE) &= DAS_2 \frac{k_1}{k_2 - k_1}[X_1 \exp(-k_1 t) - X_2 \exp(-k_2 t)], \\
S_3(t; eKE) &= DAS_3 \frac{k_2 k_1}{(k_2 - k_1)(k_3 - k_1)(k_3 - k_2)}[(k_3 - k_2)X_1 \exp(-k_1 t) \\
&\quad - (k_3 - k_1)X_2 \exp(-k_2 t) + (k_2 - k_1)X_3 \exp(-k_3 t)].
\end{aligned} \tag{S4}$$

Here, $DAS_i$ (i=1-3) are the decay associated spectra, $k_i = 1/\tau_i^{GLA}$ are the corresponding decay rates/constants and $X_i$ are originating from the convolution of the exponential decay with the Gaussian instrument response function ($S_{IRF}$) characterized with the standard deviation $\sigma_{IRF}$

$$X_i = \exp\left(\frac{\sigma_{IRF}^2}{\tau_i^{GLA}\sqrt{2}}\right)^2 \mathrm{erfc}\left(\frac{\sigma_{IRF}^2}{\tau_i^{GLA}\sqrt{2}} - \frac{t}{\sigma_{IRF}^2\sqrt{2}}\right). \tag{S5}$$

The fit is performed on two distinct data sets recorded at different days, employing common time constants and the IRF. The analysis in the main text is focused on only one of these data sets because of higher signal to noise ratio. We additionally fit a parameter accounting for a small offset in the zero-delay position ($t_0$) and obtain a shift of ~7 fs which is within the uncertainty of our time zero determination (~30 fs). Fig. S1 shows the experimental TRPES (top), the fitted TRPES (middle) and the residuals (bottom) for one of the data sets.

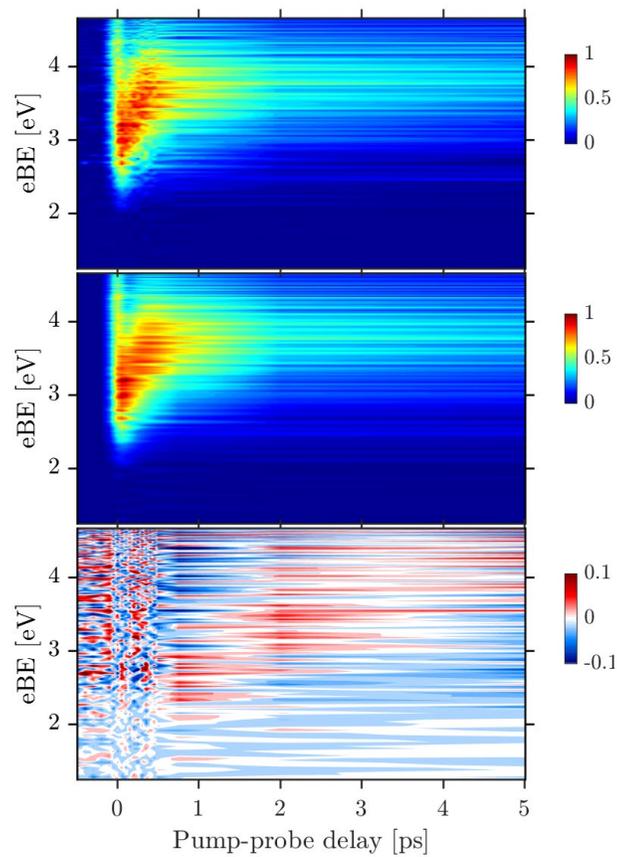

**Figure S1:** Experimental TRPES (top), the fitted TRPES (middle) and the residuals (bottom). The fitted TRPES is obtained from a GLA fit to the experimental data using an impulsive Gaussian component and three exponentially decaying components. The fit was performed on two data sets simultaneously, leading to four characteristic time constants common to both datasets.

## S3: Fitting of electron binding energy spectra

Figs. S2 and S3 show the experimental eBE spectra (circles) at different pump-probe delays $t$ without and with subtraction of the impulsive component, respectively. The black full lines are fits to the experimental spectra using an exponentially modified Gaussian function:

$$I(\text{eBE}) = \frac{a\lambda}{2} \exp\left(\mu\lambda + \left(\frac{\lambda\sigma}{\sqrt{2}}\right)^2 - \lambda \cdot \text{eBE}\right) \text{erfc}\left(\frac{\mu + \lambda\sigma^2 - \text{eBE}}{\sigma\sqrt{2}}\right), \qquad (S6)$$

$I(\text{eBE})$ is the electron signal at binding energy eBE, $\mu$ is the center and $\sigma$ is the standard deviation of the Gaussian distribution, $\lambda$ describes the asymmetric tail and $a$ is the amplitude. The vertical electron binding energy (VBE, red dashed lines) is determined as the position at the maximum of $I(\text{eBE})$ and the binding energy at 50% of the maximum signal (HBE, blue dashed-dotted lines) is also shown. The comparison of Figs. S2 and S3 reveals that both methods provide very similar VBEs and inflection points at $t > 60$ fs.

Usually, simple Gaussian functions are used to fit the eBE spectra. Fig. S4 shows the corresponding fits with a Gaussian function. The comparison of Figs. S3 and S4 illustrates that Gaussian functions are a reasonable choice for longer time delays, but that they provide somewhat less good fits (in the low eBE region) at shorter times (e.g. 60 – 200 fs) compared with the exponentially modified Gaussian function. The comparison of Figs. S3 and S4 illustrates that the fitting method does not significantly influence the values of the VBEs and the HBEs.

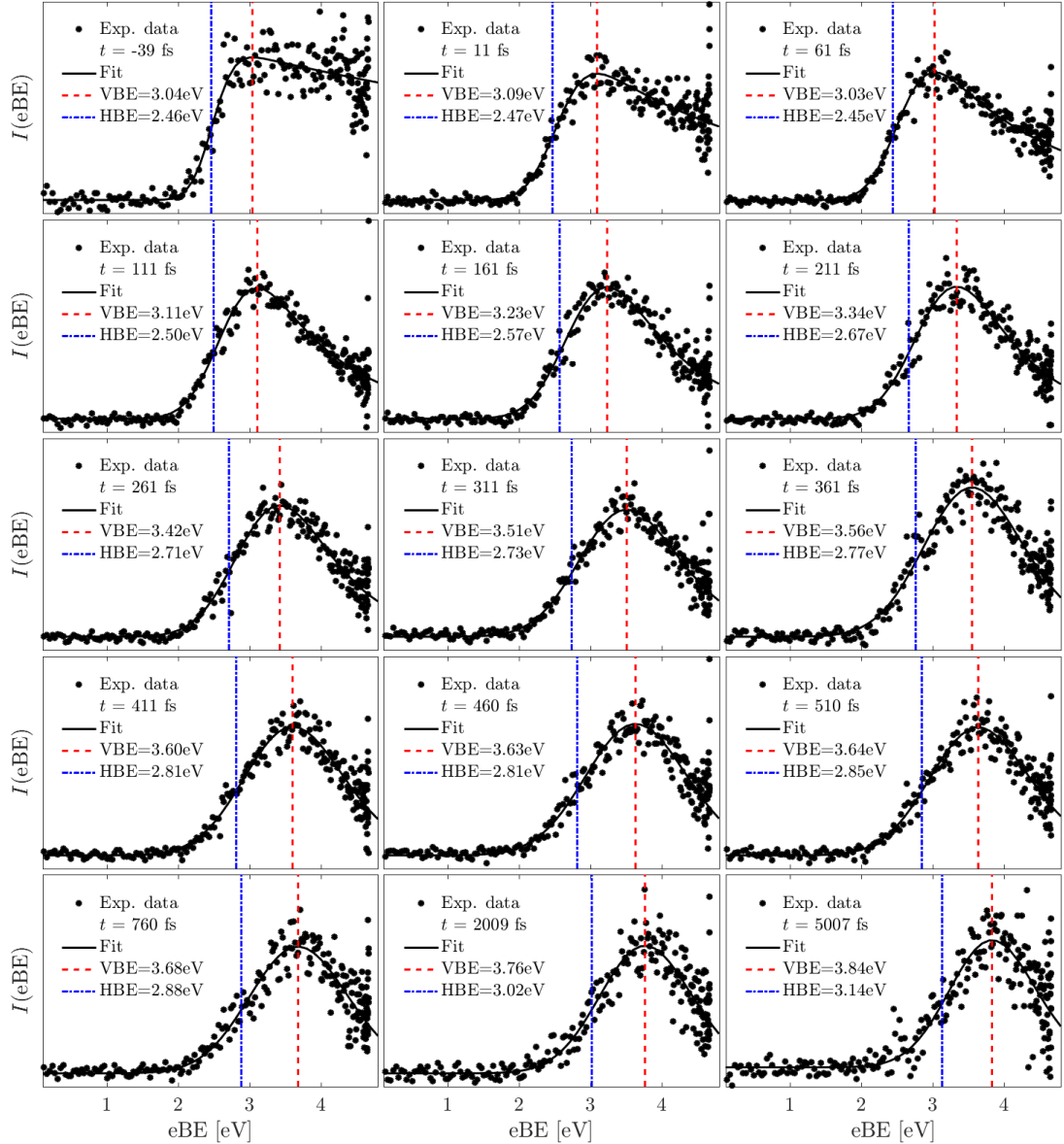

**Figure S2:** Circles: Experimental electron binding energy spectra (without subtraction of the impulsive component). Black full line: Fit with an exponentially modified Gaussian function (Eq. S6). *t* is the pump-probe delay. Red dashed line: Position of the VBE. Blue dashed-dotted line: Position of the binding energy at half maximum (HBE).

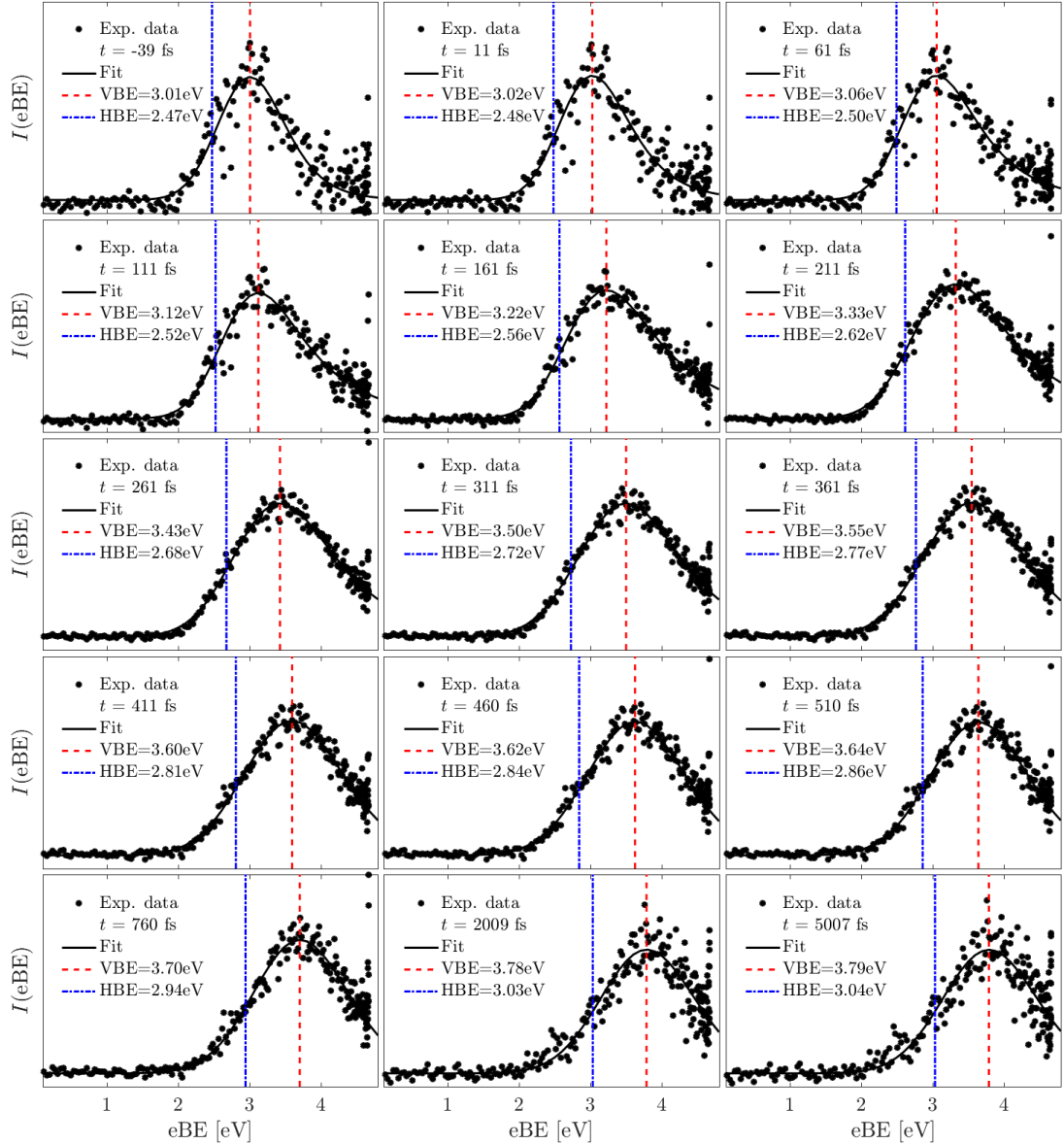

**Figure S3:** Circles: Experimental electron binding energy spectra obtained after subtraction of the impulsive component obtained from GLA. Black full line: Fit with an exponentially modified Gaussian function (Eq. S6). $t$ is the pump-probe delay. Red dashed line: Position of the VBE. Blue dashed-dotted line: Position of the binding energy at half maximum (HBE).

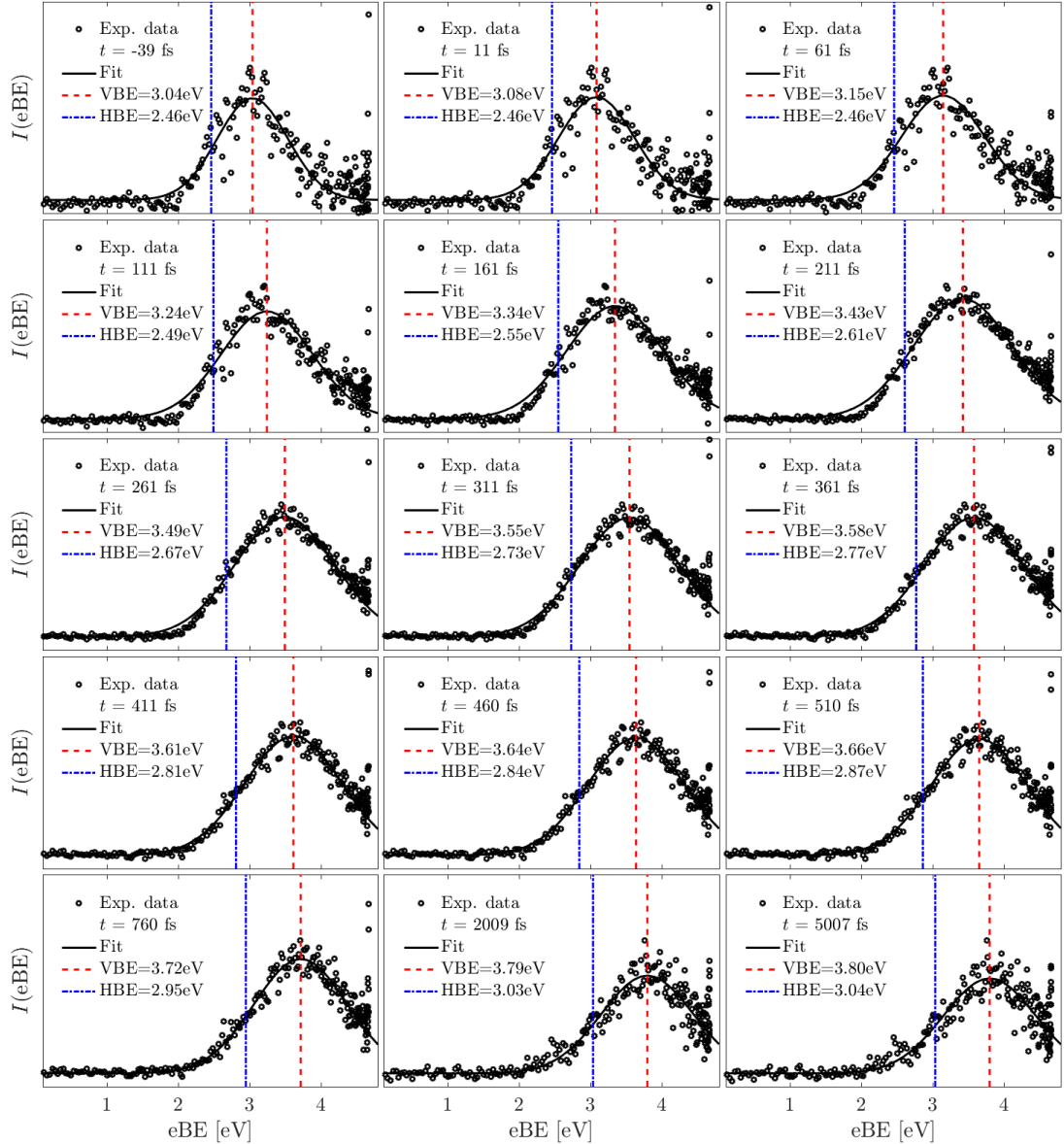

**Figure S4:** Circles: Experimental electron binding energy spectra after subtraction of the impulsive component obtained from GLA. Black full line: Fit with Gaussian functions. *t* is the pump-probe delay. Red dashed line: Position of the VBE. Blue dashed-dotted line: Position of the binding energy at half maximum (HBE).